\documentstyle[aps,prl,epsfig,amsmath,multicol]{revtex} 	

\newcommand{\be}{\begin{eqnarray}}
\newcommand{\ee}{\end{eqnarray}}

\begin{document}

\title{Defect Structures in the Growth Kinetics of the 
Swift-Hohenberg Model}
\author{Hai Qian and Gene F. Mazenko}
\address{James Franck Institute and Department of
Physics, University of Chicago, Chicago, Illinois 60637} 
\date{Oct. 18, 2002}
\maketitle
\begin{abstract}
The growth of striped order resulting from a quench of
the two-dimensional Swift-Hohenberg model is studied in the
regime of a small control parameter and quenches to zero
temperature. We introduce an algorithm for finding and identifying the disordering
defects (dislocations, disclinations and grain boundaries) at a given 
time. We can  track 
their trajectories separately.
We find that the coarsening of the defects and lowering of the effective
free energy in the system are governed by a growth law $L(t)\approx t^{x}$
with an exponent $x$ near $1/3$.  We obtain scaling for the correlations
of the nematic order parameter with the same growth law.  The scaling for the
order parameter structure factor is governed, as found by others, by
a growth law with an exponent smaller than $x$ and near to $1/4$. By comparing two  systems with
different sizes, we clarify the finite size effect. We find that the
system has a very low density of disclinations compared to that for
dislocations and fraction of points in grain boundaries. We also measure
the speed distributions of the defects at different times and find that
they all have power-law tails and the average speed decreases as a power law.
\end{abstract}
\draft
\pacs{PACS numbers: 05.70.Ln, 64.60.Cn, 64.75.+g, 98.80.Cq}

\begin{multicols}{2}

\section{Introduction}

What are the defects which control the long-time
ordering of systems growing a striped pattern?
This question arises in a variety of physical
contexts\cite{BN}.  Here we are motivated by 
the recent experiments\cite{Harrison,H2002} 
investigating the ordering of a two dimensional
diblock copolymer system.
The system studied offers a physical realization of
the ordering in an isotropic two-dimensional smectic material.
In these experiments they
found that the late-time ordering 
satisfies scaling with a growth law $L\approx t^{x}$ with $x=1/4$ 
and the final stages of ordering are governed by the
annihilation of sets of disclination quadrapoles.
In this paper we address the question:  
Is the ordering in this physical system described by the
Swift-Hohenberg (SH) model\cite{SH}, the simplest model
one can construct to govern the ordering in
stripe forming systems?

We investigate the growth kinetics of the Swift-Hohenberg 
model for a small control parameter ($\epsilon=0.1$)
in two dimensions and quenches to zero temperature.  
It is this regime which appears most likely to correspond
to the experimental situation. In large $\epsilon$ regime the system
evolves to a glassy state.
We focus primarily on the
defect structures generated in the ordering of the
system.  In the most naive picture
of this ordering process one can think in terms of an 
initial local layering,
as in a smectic, in some direction.  This ordering can be
disrupted by point defects: dislocations and disclinations.
This suggests a coarsening picture with annihilating point defects
similar to the case of the XY model\cite{BRAY} and a
growth law with exponent $x=1/2$.  This simple picture is not seen 
in simulations.
We find, in agreement with the numerical results
of Hou \textit{et al.} \cite{HSG}
and Boyer and Vi\~nals \cite{BVdc}, 
that the
defect structures for the SH model are dominated by grain boundaries which
persist for long times.  Unlike the case of an XY model, the ordering is not
dominated by annihilation of isolated point defects.  These are
observed but are not the dominant structures.

We find numerically,
at late times after finite size effects enter, that the system becomes
anisotropic, and the grain boundaries shrink.  In this case
one sees a cross-over
to an effective growth exponent $x=1/2$. 

We give below a detailed numerical study of the 
statistical properties of the defects disrupting  
striped pattern formation
in the SH model.
In order to carefully discuss the defects we need a reliable
filter for finding them.  We present an algorithm which effectively
locates defects and grain boundaries for any control parameter $\epsilon$.
We can distinguish between
grain boundaries and other defects, and track their trajectories separately.
We compare this method to the other approaches used in earlier
work in appendix A.

There are a number of ways of characterizing the degree of ordering in
these systems: (i)  Counting the number and size of defects and their
evolution with time. (ii)  Monitoring the lowering of the average
effective driving free energy as a function of time.  (iii)  Evaluation
of the nematic order parameter correlation function and its associated
scaling behavior.  (iv)  Evaluation
of the order parameter structure factor and its associated
scaling behavior.  We find that (i), (ii), and (iii) can all
be characterized by a single growth law with the exponent near $1/3$, while
the order parameter scaling, as found by others, is characterized 
by a growth law with the exponent near $1/5$.

\section{Swift-Hohenberg Model}

The Swift-Hohenberg model for a scalar order parameter, $\psi$,
is specified by the equation of motion
\be
\frac{\partial \psi}{\partial t}
=\epsilon \psi -\psi^{3}-\left(q_{0}^{2}+\nabla^{2}\right)^{2}\psi+\zeta
\ ,
\ee
where $\epsilon$ is a positive control parameter,
$q_{0}$ is the magnitude of an ordering wavenumber and $\zeta$ is the
Gaussian noise satisfying $\langle\zeta({\bf r},t)\zeta({\bf
r}',t')\rangle=2\Gamma\delta({\bf r}-{\bf r}')\delta(t-t')$, where the
noise strength $\Gamma$ is proportional to the final temperature
governing the system after a quench. We
will focus here on quenches to zero temperature where we can set
$\Gamma$ and the noise $\zeta$ to zero.
We are interested in the growth kinetics problem where we prepare
this system initially in a completely disordered state.  We then allow
the system to evolve forward in time to form a striped pattern. For 
example one
could choose
\be
\langle \psi ({\bf x},t_{0})\psi ({\bf y},t_{0})\rangle
=\Psi_{0}^{2}\,\delta\left({\bf x}- {\bf y}\right) \ ,
\ee
where $\Psi_0^2$ is a constant. However the precise form of the initial
conditions is not important \cite{BRAY}.

This model can be formulated as a Langevin equation driven
by an effective Hamiltonian:
\be
{\cal H}_{E}=\int d^{2}x\left\{- \frac{\epsilon}{2}\psi^{2}
+\frac{1}{4}\psi^{4}
+\frac{1}{2}\left[\left(q_{0}^{2}+\nabla^{2}\right)\psi\right]^{2}\right\}
\label{eq:3} \ .
\ee
If we introduce
\be
E(t)=\langle {\cal H}_{E}\rangle_{t}\ ,
\ee
where the average is over an ensemble of initial conditions,
then $E(t)$ is lowered as
the system orders in a striped pattern
with wavenumber $q_{0}$.  

Eventually
the system approaches an ordered
state described approximately by the single-mode approximation \cite{PM79}
where, assuming layering along the z-direction,
\be
\psi_0 = A_0\, \cos q_{0}z \ .
\ee
If we put this ansatz into Eq.(\ref{eq:3}), assume that the system is an 
integral
number of wavelengths in the z-direction, and minimize with respect to the
amplitude $A_0$, we obtain the results,
\begin{eqnarray}\displaystyle
A_{0}^{2}&=&\frac{4\epsilon}{3} \ ,\\
\langle \,\psi_0^2 \,\rangle&=& \frac{q_0}{2\pi}\int_0^{2\pi/q_0}(A_0\, \cos
q_0z)^2\,dz = \frac{2 \epsilon}{3}\ ,
\end{eqnarray}
and
\be
E_{eq}=-\frac{\epsilon^{2}}{6}S\ ,
\ee
where $S$ is the area of the system. Pomeau and Manneville\cite{PM79} 
have shown
that this is a very good approximation for the ``ground'' state even for
moderately large values of $\epsilon$. In the growth kinetics context 
the approach to
equilibrium is monitored by
\be
\Delta E(t)\equiv E(t)-E_{eq}\propto L_{E}^{-1}(t) \ ,
\ee
and
\be
\Delta \psi^{2}(t)\equiv
\langle\,\psi_0^2\,\rangle-\langle \psi^{2}\rangle_{t}\propto
L_{\psi}^{-1}(t) \ ,
\ee
where $L_E(t) \propto L_{\psi}(t)$ \cite{NT}.

Another measure of the ordering in the system is given by
considering the director field
\be
\hat{n}({\bf x})=\frac{\nabla \psi ({\bf x})}{|\nabla \psi ({\bf x})|}\ ,
\ee
and the associated nematic order parameter
\be
Q_{\alpha\beta}=Q_0\,\left[\hat{n}_{\alpha}\hat{n}_{\beta}-\frac{1}{2}\,\delta_{\alpha\beta}\right]
\ .
\label{eq:12}
\ee
In two dimensions, however, all of the information in this
order parameter is contained in the quantity $\cos 2\theta$
where $\hat{n}=(\cos \theta , \sin \theta)$. It is
easy to show, for example, that
\begin{eqnarray}
&&\,C_{nn}({\bf x},{\bf y},t)
\equiv 2\langle \,\mathrm{Tr}\, Q({\bf x},t)Q({\bf y},t)\,\rangle_{t}
\nonumber \\
&&=
\left\langle\, \cos\left[ (\varphi ({\bf x},t)-\varphi ({\bf
y},t)\right]\,\right\rangle_{t} \ .
\end{eqnarray}
where
\be
\varphi ({\bf x},t)=2\theta ({\bf x},t)
~~~.
\ee
If we define
\be
\hat{B}_{x}=\hat{n}_{x}^{2}-\hat{n}_{y}^{2}
\label{eq:15}
\ee
\be
\hat{B}_{y}=2\hat{n}_{x}\hat{n}_{y}
\label{eq:16}
\ee
then
\be
C_{nn}({\bf x},{\bf y},t)=\langle \hat{B}({\bf x},t)
\cdot \hat{B}({\bf y},t)\rangle_{t}
~~~.
\ee

The nematic order parameter correlation function, $C_{nn}$,
was shown by Christensen and Bray \cite{CB} to obey scaling 
in the conventional form
\be
C_{nn}({\bf r},t)=F\left(r/L_{n}(t)\right) \ ,
\ee
where ${\bf r}={\bf x}-{\bf y}$.
Elder, Vi\~nals and Grant \cite{EVG} showed that
the scaling of the order parameter structure factor 
\be
S(k,t)=\langle \,\left|\psi_{{\bf k}}(t)\right|^{2}\rangle
=L_{s}(t)F_1((k-q_{0})L_{s}(t)) \ ,
\label{eq:19}
\ee
differs from that observed
in ordering system without stripes: $S(k,t)=L^{2}(t)F_2(kL(t))$.

\section{Review of Previous Work}

The early work on this problem focused on 
establishing the final equilibrium state reached after a 
quench.
This is a
two dimensional system and by forming stripes one has a broken
continuous symmetry. The behavior of the system at non-zero
temperatures, as for the two dimensional X-Y model, requires, as pointed
out by Toner and Nelson \cite{TN81} a treatment of both long wavelength
fluctuations in the layers and free defects. Above a Kosterlitz-Thouless
type transition one has an isotropic phase while below this transition
one has a phase
with persistent orientational  order.

In an early paper,
Elder, Vi\~nals, and Grant\cite{EVG}
carried out  a numerical analysis leading to the scaling 
solution given by Eq.(19).
Working with fixed $\epsilon=0.25$ they looked at the system's
ordering  as a function
of noise strength $\Gamma$.  They found a qualitative 
difference between low noise and high noise. For the large noise case 
they found a rapid 
(exponential) relaxation to the asymptotic stationary state  and a
power-law approach  for the lower noise case.
Their results are in agreement with the picture due to
Toner and Nelson that one has a
transition to an isotropic state for large enough noise.  There is no real 
ordering in the isotropic state
and this is why there is exponential decay to the equilibrium state.
In the ordered state one has scaling and a power-law growth law which,
for small noise, they found to have an exponent $x_s=1/4$. They found 
a smaller exponent $x_s=1/5$
at low temperatures,  but they had less statistics and there
appeared to be "difficulty removing defects".  They argued for
a late time cross over to the {\it expected} $x=1/2$ but they did not
see this.

Cross and Meiron \cite{CM} also studied the SH model
numerically in the absence of noise.
They found a $x_s=1/4$ for $\epsilon =0.25$.  The dynamics appear to
freeze for higher $\epsilon$. 
They looked at the defect structure but in a qualitative way noting
the existence of domain walls rather than a set of isolated point
defects.
The theoretical discussion in their paper is based on the phase-field 
approximation
\be
\frac{\partial\phi}{\partial t}=\left(D_{\parallel}\nabla^{2}_{\parallel}
+D_{\perp}\nabla^{2}_{\perp}\right)\phi \ ,
\ee
which from the most naive point of view suggests a growth law with exponent
$x=1/2$. They discuss some selection mechanisms which could lead
$D_\parallel$ and $D_\perp$ to adjust themselves to zero and reduce $x$
to $1/3$ or $1/4$. They concluded that they did ``not have a good theoretical 
understanding of these results''
and suggested that the defects in the problems should be
treated explicitly.

Hou, Sasa, and Goldenfeld (HSG) \cite{HSG} confirmed previous
numerical results which showed
for $\epsilon =0.25$, $x_s=1/5$ with zero noise and $x_s=1/4$ with nonzero
noise as obtained from the structure factor scaling.  They went further
and used a simple method to identify domain walls and measure their lengths
(more about this below). They measured excess energy, $\Delta E(t)$,
and the domain wall length and found that they show the same scaling
exponents $1/4$ at zero noise and $0.3$ at non-zero noise. The energy
does go to the lowest order in $\epsilon$ value of $-\epsilon^2/6$ in
the noiseless limit. They find
``defects are indeed the driving force behind the coarsening process due
to its dominant contribution to the excess energy.'' They suggest that
the phase field approach gives the wrong exponent because it does not
include the effects of defects.
For larger $\epsilon$ (=0.75) they found much slower logarithmic
growth. The system seems to become glassy.

Christensen and Bray \cite{CB} also carried out numerical work
on the SH model for $\epsilon =0.25$
and found $x_s=1/5$ for zero noise and $x_s=1/4$ for nonzero noise.
From scaling of the {\it director} correlation function they find 
exponents are $0.25$ and $0.30$ for zero and nonzero noise.
They suggest that there is a cross over to $x=1/2$ at very long times.
The theory they developed does not include defects.

Boyer and Vi\~nals \cite{BVdc} point out 
"Near the bifurcation threshold, the evolution of disordered configurations is
dominated by grain boundary motion through a background of largely immobile 
curved
stripes".  They find for small $\epsilon$ an exponent $x=1/3$ which they 
interpret as arising from a
law of grain boundary motion \cite{BVgm}.
Elsewhere \cite{BVgbp} they also point out for larger 
values of $\epsilon$ the dynamics cross over to a frozen state
with quenches to zero temperature.  This glassy behavior is associated
with grain boundary pinning.

\section{Numerical Results for SH Model}

We present here our numerical results for the SH equation. We follow
the numerical prescriptions of Bray and Christensen \cite{CB}. 
We use the finite difference
scheme on two dimensional lattice of sizes $256\times 256$ and
$512\times 512$ with periodic boundary conditions. We set
$\epsilon=0.1$, $\Delta r = \pi/4$ and $\Delta t = 0.03$. We replace
$\partial_t \psi({\bf r},t)$ by
$\left(\psi_{ij}^{n+1}-\psi_{ij}^{n}\right)/\Delta t$, and $\nabla^2
\psi({\bf r},t)$ by 
\begin{equation}
\nabla^2\psi_{ij}=\frac{1}{(\Delta
x)^2}\left[\frac{2}{3}\sum_{NN}+\frac{1}{6}\sum_{NNN}-\frac{10}{3}\right]\psi_{
ij}
\end{equation}
where $NN$ and $NNN$ mean the nearest neighbors and next-nearest
neighbors respectively.
By choosing the  proper scale of time, space and the order parameter, we 
can set $q_0=1$. The systems have eight grid points per wavelength.  We used  uniformly distributed random initial conditions.

For the smaller $256\times 256$ systems we were able to follow the ordering
process to very late stages. Some of the independent trials proceed
to a final state where we have a set of well aligned layers.

In Fig. 1 we plot $\Delta E(t)$ and $\Delta \psi^{2}(t)$ for an
ensemble of runs on a
$256\times 256$ lattice. We note that there are two regimes
where $L_{E}$ defined by Eq.(9) is described by different 
exponents. For $t_{s}< t <t_{c}$ ($t_s \approx 300$ and $t_c \approx 9000$)
we find $x_{E} \approx 0.3$, while for
$t > t_{c}$ we find $x_{E}\approx 0.5$.  The cross over at
$t \approx t_{c}$ appears to be due to the finite size effects, as we
discuss below.
For $t > t_{c}$ the system is effectively anisotropic and we find an
effective exponent
$x_E$ near to  $1/2$. 

\begin{figure} 
\begin{center}
\includegraphics[scale=.43]{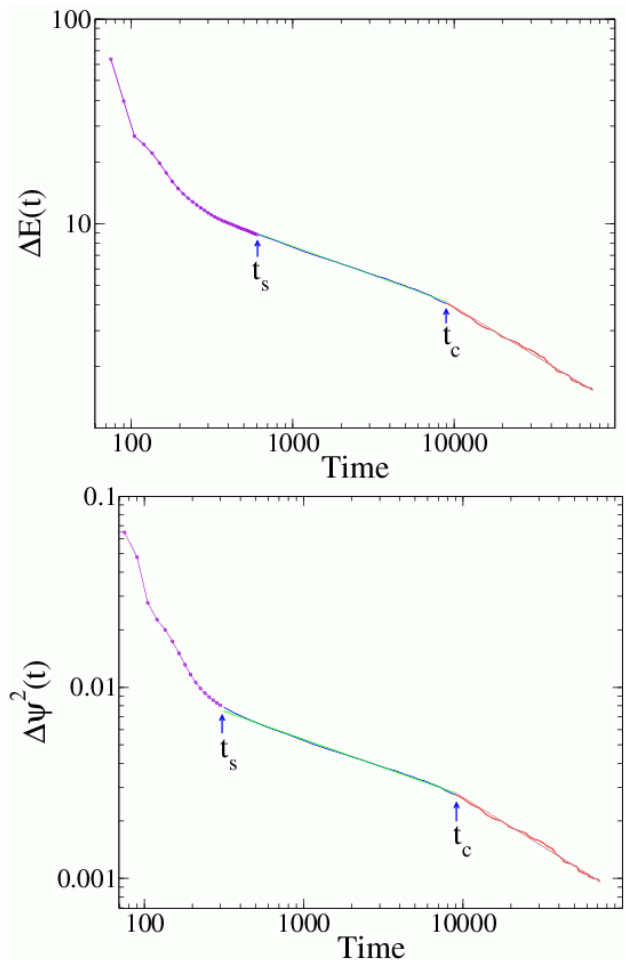}
\end{center}
\caption{\footnotesize $\Delta E(t)$ and $\Delta \psi^2(t)$ for a
$256\times 256$
system. Straight lines are used to fit different parts. Averaged over $40$ trials.}
\end{figure}

In Fig. 2 we plot $\Delta E(t)$ and $\Delta \psi^{2}(t)$
for a $512\times 512$ system.  In this case we see that
$t_{c}$ has been extended to much larger values and we have
not been able to follow the ordering process to completion.
Our fits to $\Delta E(t)$ and $\Delta \psi^{2}(t)$ in the
regime $t_{s}< t <t_{c}$ again gives, to higher accuracy,
$x_{E}=x_{\psi}=1/3$.

\begin{figure} 
\begin{center}
\includegraphics[scale=.43]{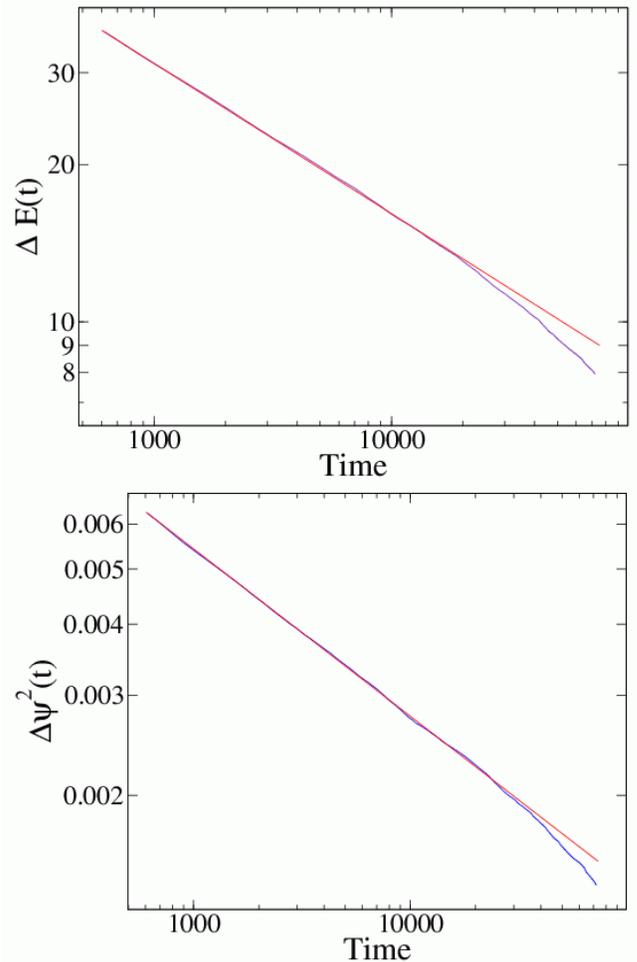}
\end{center}
\caption{\footnotesize $\Delta E(t)$ and $\Delta \psi^2(t)$ for a
$512\times 512$
system. The data for $t<t_s$ are not shown. The straight lines are used
to guide eyes. Averaged over $57$
different trials.}
\end{figure}

To probe directly the stripes' increasingly orientational order, 
we measure the {\it nematic}
order parameter correlation function $C_{nn}(\mathbf{r},t)$ 
in the $512\times 512$ system. The results,
averaged over $57$ runs, are shown in Fig. 3.  We obtain
scaling with a correlation
length obeying the growth law $L_{n} \propto
t^{0.36}$. We can estimate the time $t_c$ when the cross-over 
begins in this larger
system as follows.  The system becomes anisotropic and one
expects cross-over when the correlation length $L_n$
grows to be some substantial fraction of a lateral dimension
of the system.   In terms of ratios we can write
\begin{equation}
\frac{L_n(t_{c}(512))}{L_n(t_c(256))}\approx \frac{512}{256}=\left[\frac{t_c(51
2)}{t_c(256)}\right]^{1/3}
~~~.
\end{equation}
In the $256\times 256$ system $t_c (256)\sim 9000$, so we
obtain $t_c(512)\sim 60000$. 
Notice that in Fig. 2  the effective exponent $x_E$ begins to increase at
the time $50000\sim 70000$, which is consistent with our estimate.

\begin{figure} 
\begin{center}\includegraphics[scale=.31]{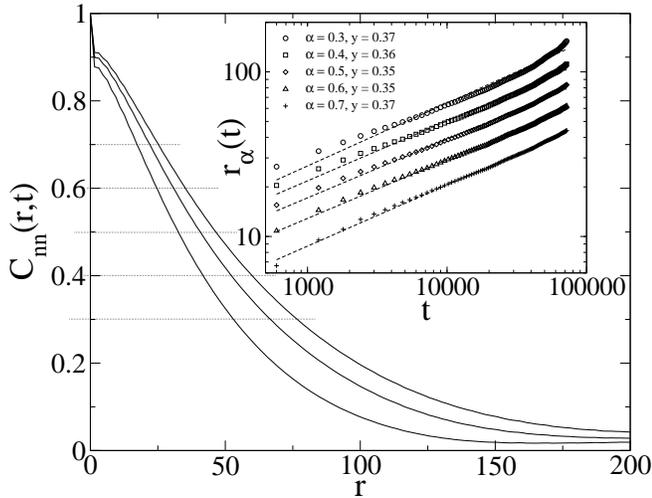}\end{center}
\caption{\footnotesize Time evolution of the correlation function 
$\displaystyle
C_{nn}(r,t)$ in
$512\times 512$ SH system
illustrated with times $6\times 10^3,1.2\times
10^4,1.8\times 10^4$ increasing from left to right. We extract the time
evolution of the correlation length $L(t)$ by monitoring the $r_\alpha
(t)$ for which $C_{nn}(r_\alpha(t))=\alpha$, where we choose
$\alpha = \{0.3,0.4,0.5,0.6,0.7\}$. The scaling exponent $x_n$ is
extracted from the log-log plot insert of $r_\alpha(t)$ v.s. t by fitting it
with a straight line. Averaged over $57$ trials.}
\end{figure}

\begin{figure} 
\begin{center}\includegraphics[scale=.31]{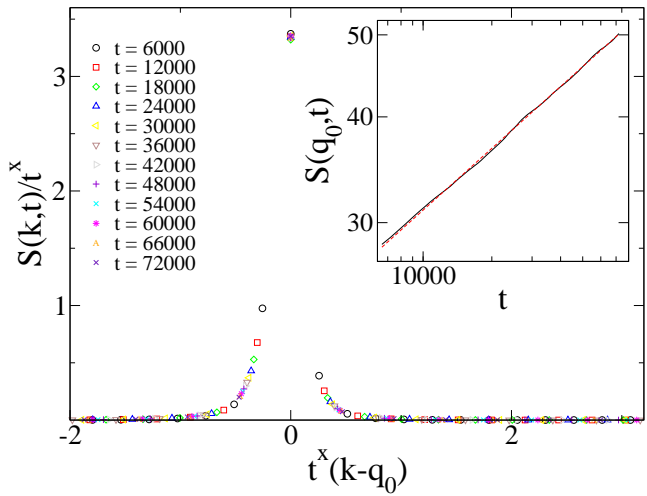}\end{center}
\caption{\footnotesize The structure factor $S(k,t)=\langle
|\psi(\mathbf{k},t)|^2 \rangle$ in the $512\times 512$
system. The log-log plot of $S(q_0,t)$ v.s. $t$ can be fit to $t^x$ with
$x=0.24$. The scaling collapse of the structure factor was obtained with
$x=0.24$ as the scaling exponent. Averaged over $57$ independent trials.}
\end{figure}

In Fig. 4 we plot the structure factor $S({\bf k},t)$ and show
that scaling holds in the form given by Eq.(\ref{eq:19}) with a growth law
characterized by an exponent $x_{s}=0.24$ as shown in the insert.
Our results here agree with those found previously that the
exponent governing the growth law for the structure factor is
significantly smaller than that governing the nematic order
parameter.

\section{Defect Structures and dynamics}

In Fig. 5 we show a typical configuration for the Swift-Hohenberg 
model for a quench to zero
temperature after a time $12000$ for a $512\times 512$ system.  Notice the
rather complicated structure which includes dislocations,
disclinations and grain (domain) boundaries. Our main focus in this
paper is to study the statistics of these defects.
In appendix A we discuss an algorithm for picking out the defects and
tracking their motions. 

\begin{figure} 
\begin{center}\includegraphics[scale=.53]{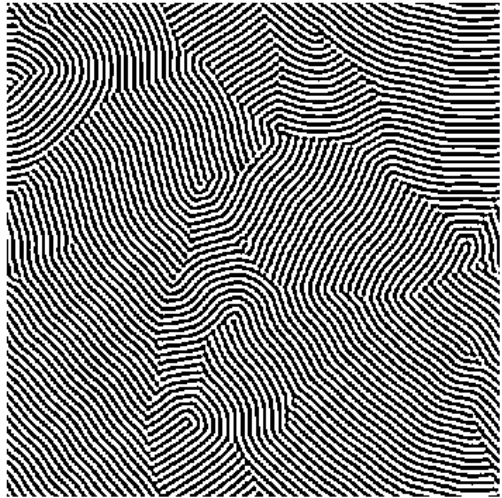}\end{center}
\caption{\footnotesize A typical configuration for the SH model for a
quench to zero temperature after $t=12000$ in a $512\times 512$ system. The 
black points correspond to $\psi({\bf x})>0$, and the
white points to $\psi({\bf x})<0$. }
\end{figure}

If we look at Fig. 5 we see that it shows a complicated situation
with a variety of different defect structures which one can
identify by eye at the length scale of several layer spacings.
At a more fundamental level we need a way of identifying which
points in space, at the level of each site on the numerical
grid, are part of a defect.
At the shortest length
scale in the problem
the order parameter is $Q_{\alpha\beta}$
 defined by Eq.(\ref{eq:12}).
For this two-dimensional system this can be replaced
by the vector order parameter ${\bf\hat{B}}$ defined by
Eqs.(\ref{eq:15}) and (\ref{eq:16}).  The assumption is that all of 
the defects in the system
can be built up from the $\pm\frac{1}{2}$ disclinations in the 
director field $\hat{n}$ which translate into vortices with charge
$\pm 1$ for the field ${\bf\hat{B}}$. 
We identify these defects by looking for the cores of the vortices.
We can find the cores of the defects by looking from those sites
where ${\bf\hat{B}}$ is changing rapidly. 
We can define
\be
A=\sum_{\alpha ,\beta}\left(\nabla_{\alpha}B_{\beta}\right)^{2}
\ee
and
identify defect points as those sites where $A$ is larger than some
value.  Notice that $A$ can also be written in the form
\be
A=4\sum_{\alpha ,\beta}\left(\nabla_{\alpha}n_{\beta}\right)^{2}
=\left(\nabla_{\alpha}\varphi\right)^{2}
~~~.
\ee
The precise numerical determination of $A$ is discussed in
appendix A.  Notice that $A$ is proportional to the
gradient energy for an isotropic nematic.

In analyzing their 
experimental data Harrison, et al. \cite{Harrison,H2002} found a set
of {\it fundamental} disclinations and from these built up
dislocations as  bound disclinations with opposite
charge.  They used this procedure to identify a large 
dislocation density. Most of the fundamental disclinations
went into forming these dislocations since in the end the
ratio of dislocations to the remaining disclinations was
about ten to one.  In our case 
the situation is complicated by the grain boundaries.  We first
separate the defects into compact point defects and larger
grain boundaries.  For the point defects we determine the
topological charge by taking the usual phase-angle path
integral around the center of mass of the defect.  Those
defects with plus or minus unit charge are identified as
disclinations, while those with zero charge are dislocations. Then we
can track the motion of each single defect.

\begin{figure} 
\begin{center}\includegraphics[scale=0.31]{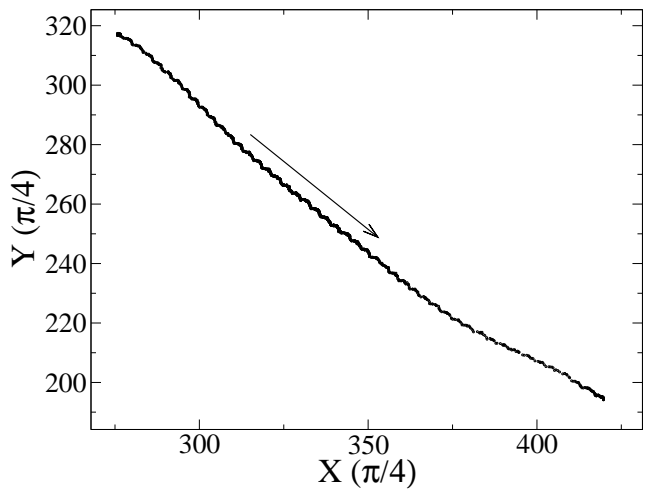}\end{center}
\caption{\footnotesize A typical example of the track of a dislocation's 
``mass center'', see appendix A. The dislocation moves along the arrow. It starts at
$t=1590$ and disappears at $t=29220$. Notice the small arcs along the
curve, the diameters of the arcs are about $2\pi$, which is equal to two layer spacing.}
\end{figure}

As an example of the method we show in Fig. 6
the path of a dislocation. We see that some dislocations travel 
over long distances during
very long times. It seems that the dislocations are more stable 
when compared to
grain boundaries and disclinations. Our simulations on $256\times 256$ system show
that after the annihilation of point defects
and grain boundaries, some dislocations still exist in the system. 
Our simulations show that there are
also dislocations which are pinned and move little. 

The number of disclinatins is quite small. And we notice that they are
rather immobile, which is consistent with  Boyer and Vi\~nals' discussion \cite{BVdc}.

The most important motion of grain boundaries is that they can move over
long distances and combine with other grain boundaries. 
As shown in Fig. 7, two grain boundaries can combine to
form a larger grain boundary. Thus the number of grain boundaries 
decreases while their average
size increases. This process happens on a time scale of the
order $1000$ dimensionless time units.

\begin{figure} 
\begin{center}
\includegraphics[scale=.28]{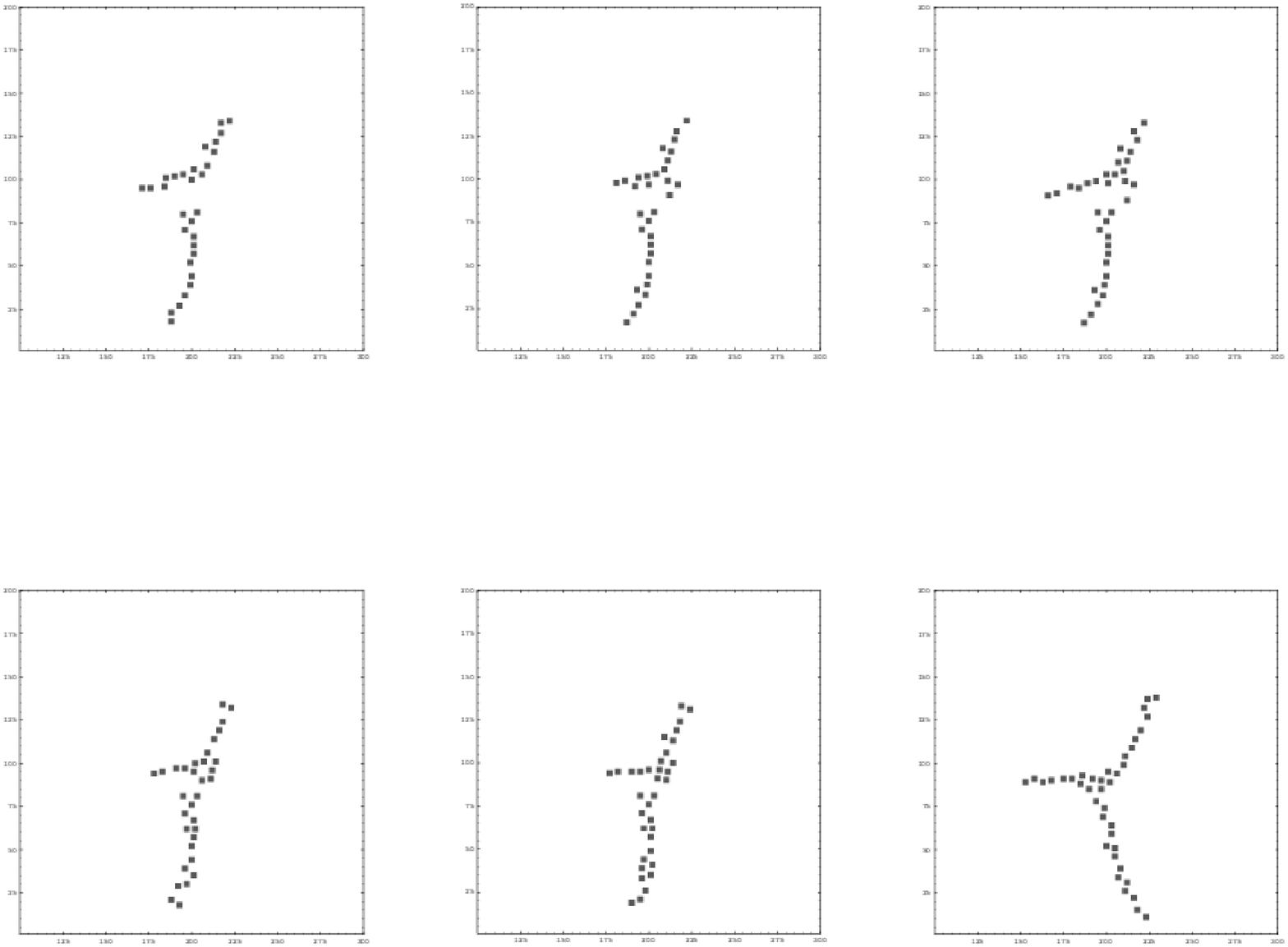}
\end{center}
\caption{\footnotesize The combination of two grain boundaries in a
$512\times 512$ system. The portion shown is $200\times
200$. From left to right and top to bottom, the
times are $t = 2880$, $2955$, $3030$, $3105$, $3180$, $4710$. Not all the
points in the grain boundaries are shown.}
\end{figure}

\begin{figure} 
\begin{center}
\includegraphics[scale=.33]{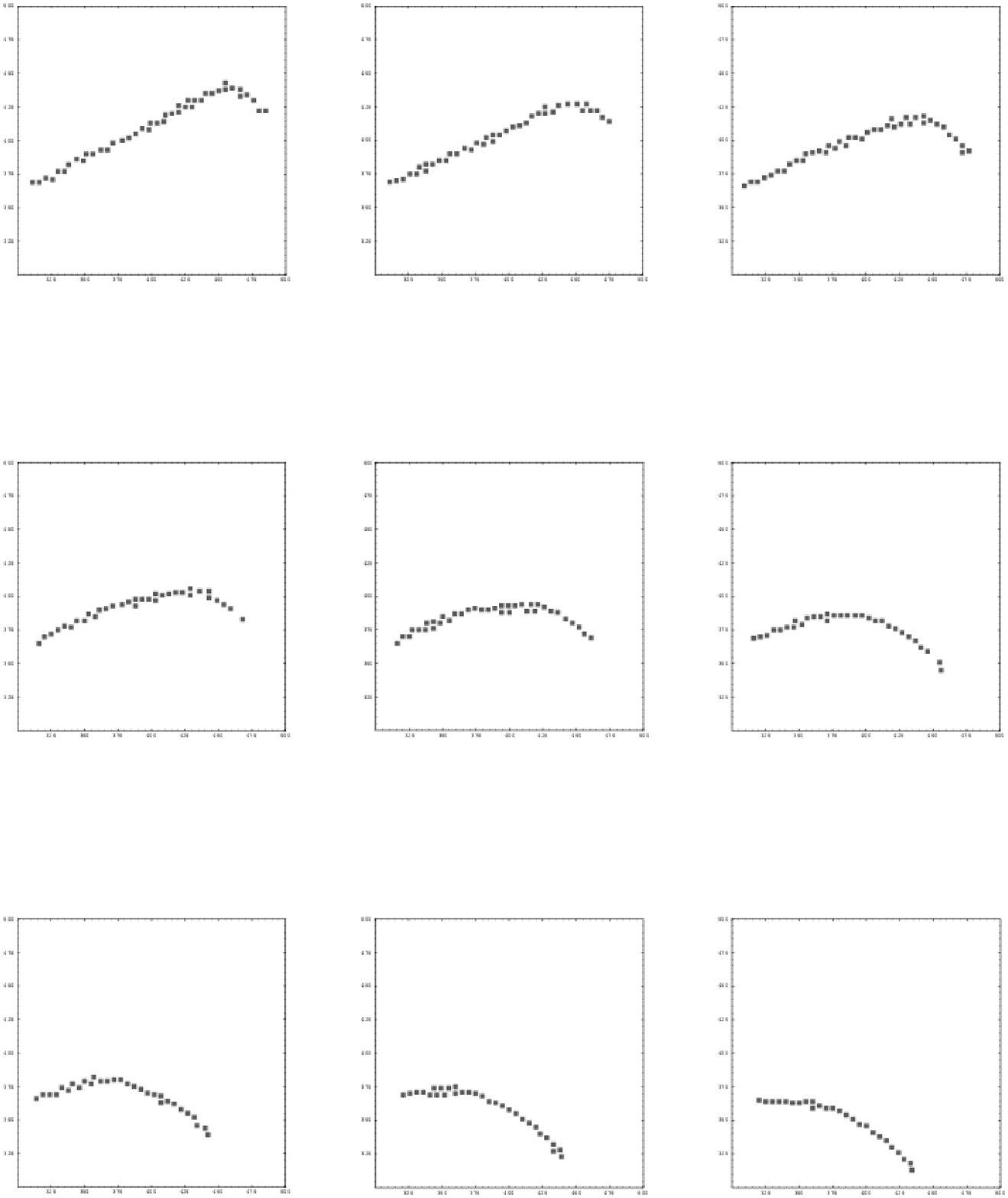}
\end{center}
\caption{\footnotesize The motion of a grain boundary in a $512\times
512$ system. The portion shown is $200\times
200$. From left to right and top to bottom, the times
are $t=11415$, $13665$, $15915$, $18165$, $20415$, $22665$, $24915$, $27165$,
 $29415$. Again not all the points in the grain boundary are shown.}
\end{figure}

As shown in Fig. 8, one grain boundary can sweep across a quite large
area. At the same time its size decreases. This process occurs on a
time scale of the order $10000$. According to our
observations, the grain boundaries' motions also relieve the stripe
curvatures through disclination annihilations. After one grain boundary
passes through a disclination, the disclination disappears. This is
consistent with Boyer and Vi\~nals' prediction \cite{BVdc}.

\begin{figure} 
\begin{center}\includegraphics[scale=.31]{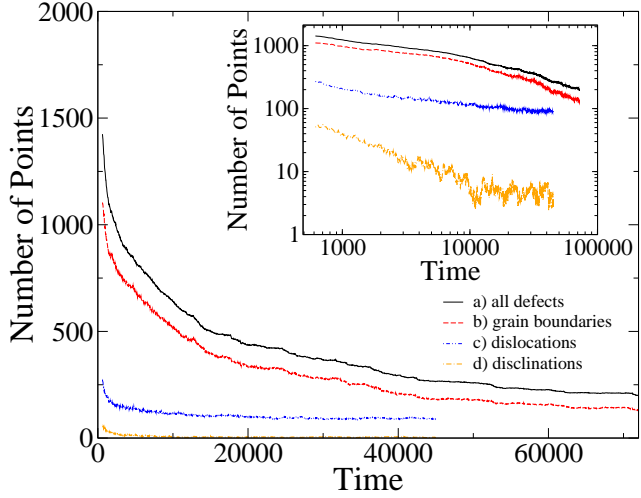}\end{center}
\caption{\footnotesize The number of points in point defects and grain
boundaries in the  $256\times 256$ system. The data for all defects and
domain walls are  averaged over $40$ trials. The others are averaged
over 38 trials.}
\end{figure} 

Next we focus on the statistics of the
defects generated by the model.
In Fig. 9 we plot the total number of points in grain
boundaries,  dislocations and disclinations separately for the $256\times 256$ 
system. 
We see that the grain boundaries dominate. In the
scaling regime ($t_s<t<t_c$), we see that the number of points
corresponding to grain boundaries and all defect points, the curves $a$
and  $b$ can be fit to $\sim 
t^{-1/3}$. At late stages the disclinations
disappear, while the dislocations and grain boundaries persist. The number of
disclinations decreases much faster than the other defects.

\begin{figure} 
\begin{center}\includegraphics[scale=.31]{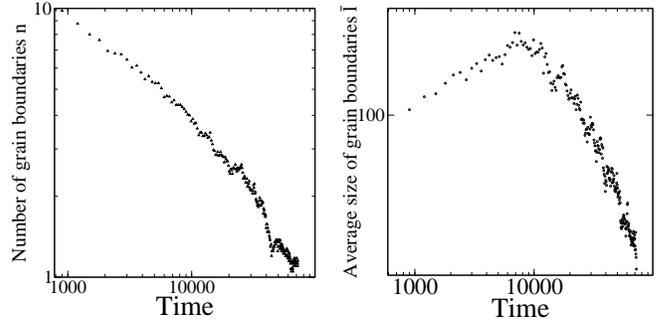}\end{center}
\caption{\footnotesize The average number of grain boundaries and the average
number of points in a grain boundary. $256\times 256$ system 
averaged over $40$ trials.}
\end{figure}

In Fig.  10 we plot the average number of grain boundaries $\bar{n}$  and the
average size of a grain boundary $\bar{l}$ for the $256\times 256$ system. We 
use the number of points in one grain boundary
as a measure of its size. For $t_s<t<t_c$, $\bar n$ decreases but $\bar l$
increases due to the combining of grain boundaries. The
shrinkage of their sizes is not as important as the combinations. However for
$t_c \sim 9000$ the correlation length $L_n$ is the same order as the
system's size, and the large grain boundaries stop growing. After that
the shrinkage is important \cite{SHRINK}. In the scaling regime ($t_s<t<t_c$), 
$\bar n
\sim t^{-0.45}$, and $\bar{l} \sim t^{0.13}$. So $\bar n \bar{l} \sim
t^{-1/3}$, which is consistent with Fig. 9.

\begin{figure} 
\begin{center}\includegraphics[scale=.31]{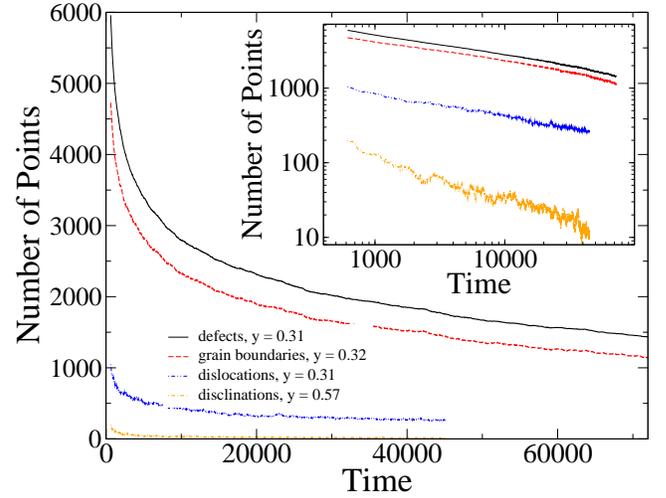}\end{center}
\caption{\footnotesize The average number of points in defects and grain
boundaries in the $512\times 512$ system. Most of the points are in grain 
boundaries. All
the curves can be fit to $t^{-y}$. The data for defects and domain
walls are averaged over $57$ trials, and the other data are averaged
over 20 trials.}
\end{figure}

\begin{figure} 
\begin{center}\includegraphics[scale=.31]{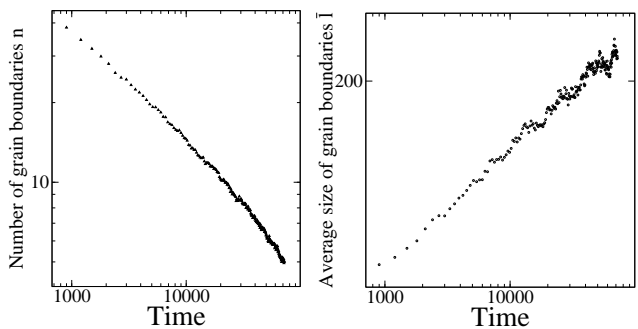}\end{center}
\caption{\footnotesize $\bar n$ and $\bar{l}$ v.s. time in the $512\times 512$ 
system. The data are averaged over $57$ independent trials.}
\end{figure}

In Fig. 11 we plot the total number of points in grain
boundaries, dislocations and disclinations separately for the $512\times 
512$ system. The number of points in disclinations is much smaller than
that of dislocations and at very late stages it decreases to the order
of 1, which in fact indicates the disappearance of disclinations. Now the
scaling regime extends to much longer times. The domain walls and
dislocations' scaling exponents are both $1/3$, which is same to the
scaling of the energy. However, the scaling
exponent of disclinations, given by 0.57 is much larger. So we conclude that 
disclinations are not the dominant structures in SH system.

In Fig.  12 we plot the number of grain boundaries $\bar{n}$ and the
average size of a grain boundary $\bar{l}$ for $512\times 512$
system. All of our data falls in the scaling regime.. 
The plot of the average
number $\bar n$ of grain boundaries versus time $t$, can be fit to $\bar n \sim
t^{-0.49}$ and the average size $\bar{l}$ v.s. time $t$, can be fit by 
$\bar{l} \sim t^{0.17}$.
So we have $\bar n \bar{l} \sim L^{-1} \sim t^{-1/3}$, which is
consistent with the results shown in Fig. 11.

Although we did not count
the number of disclinations directly, it is proportional to the number
of lattice points in disclinations,
i.e. $t^{-0.57}$. This is because the average number of lattice  points in one point
disclination, which is about $10\sim20$, is quite stable during the
simulation. By the same reasoning, we find that the number of dislocations
is proportional to  $t^{-1/3}$. It is interesting to note that the number of grain
boundaries scales as $t^{-0.49}$. This exponent is near to that for
disclinations. 

The number of grain
boundaries is about 5 at $t\sim 70000$, the number of disclinations is
about $0$ or $1$ at $t\sim 50000$,
and the number of dislocations is on the order of 10 at $t\sim 50000$, as can be seen in
Figs. 11 and 12.

\begin{figure} 
\begin{center}\includegraphics[scale=0.85]{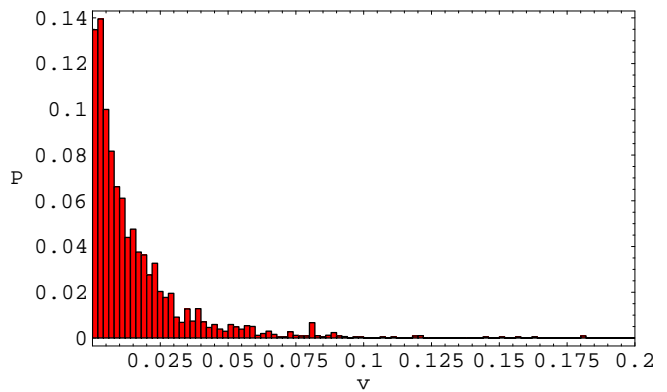}\end{center}
\caption{\footnotesize A typical example of probability density $P$
v.s. speed $v$ for the speed distribution of
dislocations at time $1350$ for the $512\times 512$ system. The distributions at other
times have approximately the same shape. Averaged over 56 trials.}
\end{figure}

Since we can track the motion of each defect, we can
measure their speeds. We define the speed of
each as the speed of its mass center (see appendix A for
the definition of ``mass center''). If in a time $\Delta \tau$, the mass center
travels over a distance $\Delta d$, then the speed is $v=\Delta
d/\Delta \tau$. If
$\Delta \tau$ is small enough, we found that $v=0$ has the biggest probability.. If $\Delta\tau$ is large enough, all the details are
coarse-grained and we observe a continuous distribution of the speed and
the largest probability appears at a non-zero speed, as is shown in
Fig. 13 where $\Delta \tau =60$. We measured the speed distributions of
domain walls and point defects separately. As we have already seen, for
point defects the number of disclinations is much smaller than that of
dislocations, so what we measure in the latter case is in fact the speed distribution of dislocations.

The speed distribution has
a long tail which decreases as a power law. The numerical fits at
different times give us different exponents. However the tail exponents
 at different times do distribute in a narrow region, as is shown in
Fig. 14. The exponents of grain boundaries are quite different from those
of point defects. If we ignore the exponents at very early times
when grain
boundaries just begin to form and at the very late times when the grain
boundaries have already disappeared, the mean value of the grain
boundaries' exponents is $-1.50$, and that of
point defects' exponents is $-2.10$. 

\begin{figure} 
\begin{center}\includegraphics[scale=0.31]{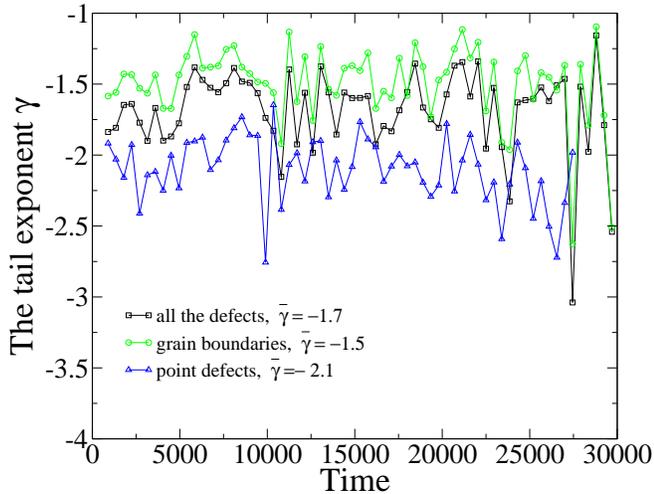}\end{center}
\caption{\footnotesize The power law exponents of the speed distribution's
tail at different times. Ignoring the data points at very early times and
very late times, the mean value of the exponents is $-1.5$ for grain
boundaries, $-2.1$ for point defects and $-1.7$ for all the defects.
Averaged over $61$
independent trials.}
\end{figure}

\begin{figure} 
\begin{center}\includegraphics[scale=0.31]{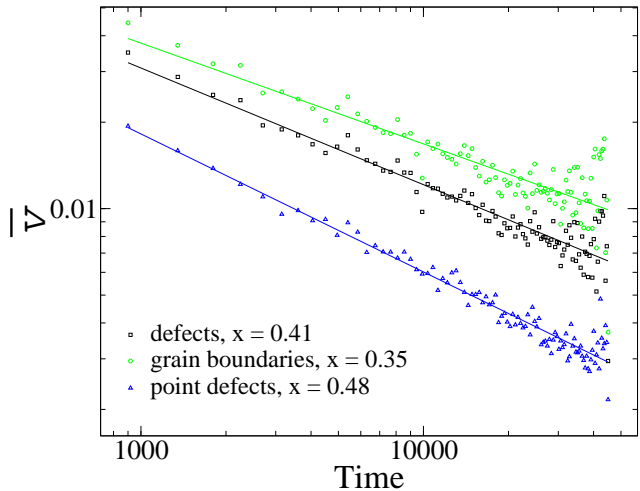}\end{center}
\caption{\footnotesize The average speed for defects and grain
boundaries. From top to bottom, the curves have the form  $v_0\times
t^{-x}$ with $v_0$ being a constant. From top to bottom, $x=0.35\pm 0.01$, $x=0.41\pm 0.01$ and 
$x=0.48\pm0.01$. Averaged over $56$
independent trials.}
\end{figure}

We
also measured the average speed of the point defects and grain
boundaries as a function of time after the quench,
as is shown in Fig. 15. The average speed of point defects decreases as  $\sim 
t^{-0.48}$;
the average speed of grain boundaries goes as  $\sim t^{-0.35}$. The
scattering of the points at late stages is due to the small data sample
at those times..


\section{Summary}

We have studied the dynamics of the defect structures in the SH model
after quenches to zero temperature with a control parameter of
$\epsilon=0.1$. We find in agreement with earlier workers that the
kinetics in the ordering regime, before finite-size effects enter,
are dominated by the existence of moving and coalescing grain
boundaries. In this regime the average size of these grain boundaries is
growing and they are relatively mobile. Under the influence of finite
size effects these grain boundaries shrink, the system becomes
anisotropic and the ordering process speeds up. 

We also measured the speed distribution of all structures that appear in
the system. The average speed is decreasing as a power law and the
distributions show a power-law behaviors at large speeds.
However we can only get a rough estimate due to the poor statistics.

Let us return to the question of whether the SH model gives a
good description of the physical system studied by  Harrison,
et al.\cite{Harrison,H2002}.  The SH model, 
for small
control parameter $\epsilon$, does give coarsening with an
exponent in roughly the same range as in the experiment
($1/4\sim 1/3$).  The ordering is constrained to be slower
then the picture where one has
a simple point defect pair annihilation process.  However,
the defect structures in the SH model and experiment
appear quite different.  The disclination quadrapole annihilations
seen in experiments are not observed in the late stages of the
evolution of the
SH model.  In the SH model grain boundaries dominate the
evolution in the scaling regime, but these structures appear
to play a limited role in the experiments.
We must make clear that our numerical results are for systems
with many fewer roll periods compared to the experimental systems
($10^{2}$ compared to $10^{5}$), so it is possible that things
change as we increase the size of the ordering system. However, our
study of 256 and 512 systems shows that they differ only in 
the time when the finite-size effect enters. This indicates that
a even larger system will display the same behavior except that the
finite-size effect enters at a even later time. 
So we conclude that the SH model does not
give a physically faithful description of the ordering in the 
experimental system.
This raises the provocative question:
Are there many different types of scenarios for ordering striped
systems?  We will address this question by looking at other
competing models for striped formation elsewhere.

\vspace{5mm}

Acknowledgments:  We thank Dr. C. Harrison for providing
us with a copy of his thesis. This work was supported by the National
Science Foundation under Contract NO. DMR-0099324.


\appendix

\section{The algorithm for picking out defects and finding
    domain walls}

Hou, \textit{et al.} \cite{HSG} proposed a method, the HSG method, to measure 
the length of grain boundaries.
They computed the quantity $A^2 \equiv \psi^2+(\nabla \psi)^2/q_0^2$, and if
the calculated $A^2$ is bigger than an upper filter $0.7\times 
Max(A^2)+0.3\times
AVG(A^2)$ or smaller than a lower filter $0.7  \times Min(A^2)+0.3\times 
AVG(A^2)$,
that point is counted as belonging to a domain wall. When $\epsilon$ is
small, this method gives quite good results.

However, when $\epsilon$ increases, the original filters are no longer
applicable. They fail to pick out most of the points and the filters
must be re-chosen.
For example, at $\epsilon=0.6$, the filters $0.5\times Max(A^2)+0.5\times
AVG(A^2)$ and  $0.5  \times Min(A^2)+0.5\times AVG(A^2)$ can give a
satisfying result; while for $\epsilon = 0.75$,  $0.4\times Max(A^2)+0.6\times
AVG(A^2)$ and  $0.4  \times Min(A^2)+0.6\times AVG(A^2)$ are the better
choices. Sometimes this method is unable to pick out all the  defects for
any choice of filter.

We introduce here a 
method which works for all $\epsilon$ and picks
out all of the defects and nothing more. 

First let us define some useful quantities. Suppose the system is
 discrete on the  $x\-y$ plane, with ${\bf x}=(i,j)$ (square lattice). At a 
fixed time,
starting from the order
 parameter field $\psi ({\bf x})=\psi_{i,j}$, we can define a director
 field $\hat{n}({\bf x})$
 as given by Eq.(11) 
 where $\nabla \psi({\bf x})$ is defined by the usual finite difference
 scheme, i.e.
 \begin{eqnarray}
 \nabla \psi({\bf x})= \left(\frac{\psi_{i+1,j}-\psi_{i-1,j}}{2\,\Delta
 r},\frac{\psi_{i,j+1}-\psi_{i,j-1}}{2\,\Delta r}\right) \ ,
 \end{eqnarray}
 where $\Delta r$ is the lattice space of the system.
 In two-dimensional cases the nematic order parameter,
$Q_{\alpha \beta}$ is completely specified by the angle
 \begin{equation}
 \varphi ({\bf x}) = 2\, \theta ({\bf x})\ ,
 \end{equation}
 where
 \begin{equation}
 \theta ({\bf x}) = \arctan
\left( \frac{\hat{n}_y({\bf x})}{\hat{n}_x({\bf x})}\right)\ .
 \end{equation}

Rather than using $\varphi ({\bf x})$ given by the two equations
above we introduce some local smoothing.
First we compute
\begin{eqnarray}
&&\hat{B}_{y}= \sin \varphi({\bf x})=2 \hat{n}_x({\bf x}) \hat{n}_y({\bf x})\ ,
\nonumber \\
&& \hat{B}_{x}=\cos \varphi({\bf x})=2\hat{n}_x({\bf x})^2-1\ .
\end{eqnarray}
Then we smooth these two fields using the iterative process:
\begin{equation}
f_{(n+1)}(i,j)=\frac{1}{2}f_{(n)}(i,j)
+\frac{1}{8}\sum_{(i',j')\in NN}f_{(n)}(i',j')\ ,
\end{equation}
where $f_{(n)}$ is $\sin\varphi$ or $\cos \varphi$ after $n$ iterations,
and $NN$ means the $4$ nearest neighbors of $(i,j)$ on
the square lattice. This process will suppress the small fluctuations
of $\varphi({\bf x})$ away from the defects, while the variation of
of $\varphi({\bf x})$ near a defect core remains large. Our calculations
show that $5$ iterations provides a sufficiently smooth set of
fields for our purposes.
In the next step, we calculate
$\varphi({\bf x})$ from  $\sin \varphi({\bf x})$ and  $\cos
\varphi({\bf x})$ using
\begin{equation}
\varphi({\bf x})=\arctan \left[\frac{\sin
\varphi({\bf x})}{\cos\varphi({\bf x})}\right]\ ,
\end{equation}
where we adopt the convention that $-\pi < \varphi({\bf x}) <\pi$.

In picking a filter we want to look at the spatial variation
of the $\varphi({\bf x})$ field. 
$\nabla \varphi({\bf x})$ can be evaluated as for $\nabla
\psi({\bf x})$. However, there is a subtlety here. For example, if
$\varphi_{i+1,j}=\pi-\delta \phi_1$ and $\varphi_{i-1,j}=-\pi+\delta
\phi_2$, where $\delta \phi_1$ and $\delta \phi_2$ are small angles, 
then the difference between the nematic tensor
$Q_{\alpha\beta}(i+1,j)$ and $Q_{\alpha\beta}(i-1,j)$ should be a small
quantity. But
$\displaystyle(\varphi_{i+1,j}-\varphi_{i-1,j})/2
=\pi-(\delta\phi_1+\delta\phi_2)/2 \sim \pi$, which
means that if we calculate the change rate of $\varphi({\bf x})$ in
exactly the same way as Eq.(A1), we will get a wrong answer in this 
context. To avoid such a
problem, we define the difference between $\varphi({\bf x})$ and
$\varphi({\bf x'})$ as the
quantity with the smallest absolute value among the choices
$\varphi({\bf x}')-\varphi({\bf x})$ and
$\varphi({\bf x}')-\varphi({\bf x})\pm 2\pi$. And we use this quantity
in determining
\be
A({\bf x})=|\nabla
\varphi({\bf x}) |^2 
\ee
which is the key quantity in our analysis.

\begin{figure} 
\begin{center}\includegraphics[scale=.4]{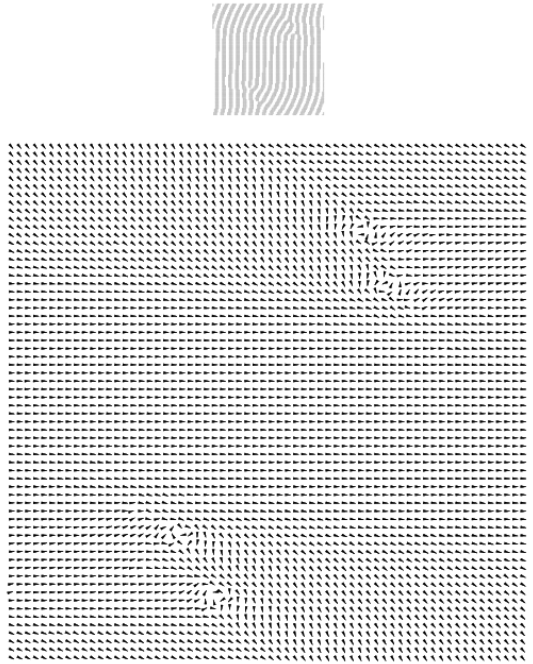}\end{center}
\caption{\footnotesize In the lower graph, the vector field $(\cos
\varphi(\vec{x}),\sin \varphi(\vec{x}))$ for the order parameter field
shown above. The
components of the vector field have been smoothed over $5$ iterations. The
lattice spacing is $\pi/4$, which means there are $8$ points in one
period of the layers. Not all the vectors on the lattice are shown.}
\end{figure}

Our method
is based on the observation that at the core region of a defect
(dislocation, disclination or part of a domain wall) the angle
field $\varphi({\bf x})$ changes
rapidly, while in the region away from the defect's core the
$\varphi({\bf x})$ field is
rather smooth, as can be seen in Fig. 16. Thus we can conclude with
confidence that those points with larger change rates of
$\varphi({\bf x})$ must belong to some defect's core region or a part
of a grain boundary.

\begin{figure} 
\begin{center}\includegraphics[scale=0.45]{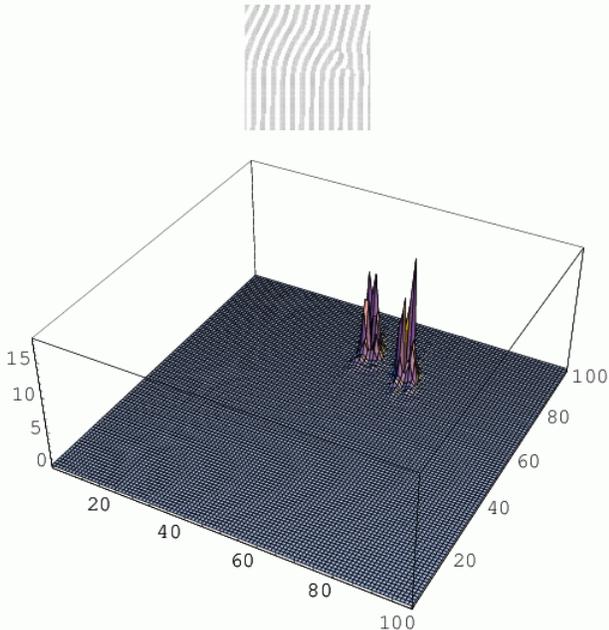}\end{center}
\caption{\footnotesize In the lower graph, the scalar field 
$A(\vec{x})$ is plotted. This corresponds to 
 the order parameter field shown  above. $A(\vec{x})$ is sharply peaked at the core regions of the defects.}
\end{figure}

We find that $A({\bf x})\approx 0$ away from defects,
but increases very rapidly in the vicinity of any defect. An example
is given in Fig. 17. Therefore as long as $A({\bf x})$ is large
enough, we
can identify the point ${\bf x}=(i,j)$ as part of the core region of a
defect. Naturally we set up a threshold $A_0$, and any point with
$4(\Delta r)^2\cdot A({\bf x})>A_0$ is counted as belonging to some defect's core. Because the
value of $A({\bf x})$ is much larger in the defects' cores than at any other 
places, a range of values of $A_0$ can be used to find  the positions of the defects core
regions. With a smaller threshold the program will pick out more points 
in the core regions,
and with a larger one it will pick out fewer points in the core regions. Our
experience shows that if $A_0$ takes the value of $2\sim 10$, the
program picks out the same defects cores and grain boundaries. 
As is shown in Fig. 18 and Fig. 19, it
picks out all the defects without  irrelevant points.

\begin{figure} 
\begin{center}\includegraphics[scale=0.63]{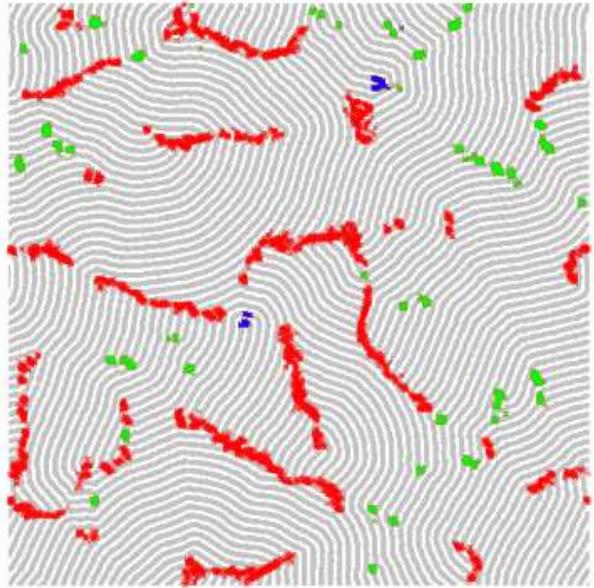}\end{center}
\caption{\footnotesize Identification of  all the defects in a $512\times 
512$
system ($\epsilon = 0.1$) with a threshold $A_0=3.5$. At each
defect core region, the $A$ field for many
points exceeds the threshold. The red points belong to domain walls, 
the green ones belong to dislocations and the blue ones belong to disclinations.}
\end{figure}

\begin{figure} 
\begin{center}\includegraphics[scale=0.63]{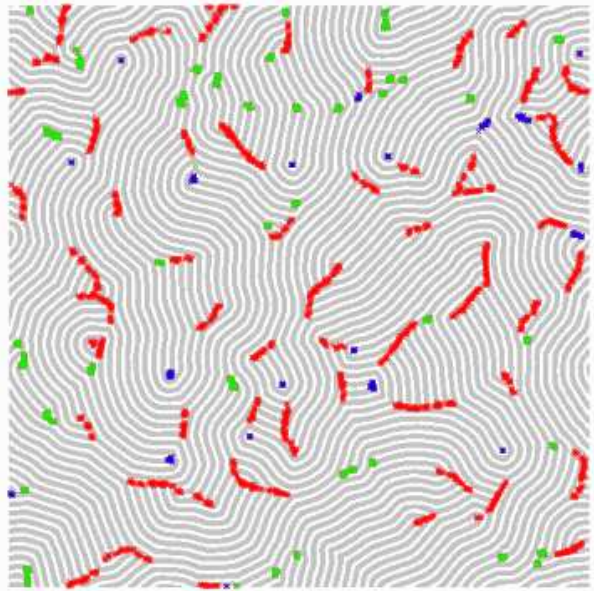}\end{center}
\caption{\footnotesize Identification of all the defects in a $512\times 
512$
system ($\epsilon = 0.5$) again with a threshold $A_0=3.5$.  Apparently the defect's density in this system is greater than
the density in Fig. 18. The domain walls are much smaller for the system
with larger $\epsilon$. There
are no domain walls for $\epsilon > 0.6$.}
\end{figure}

After we have used the above algorithm to pick out the points in the core 
regions
of the point defects and grain boundaries, we can distinguish between these
two structures. The  difference between them is obvious. The point defects are 
compact in space
while the grain boundaries are ramified. 

First, we must group the points we have identified according to whether
or not they are in the same structure. The points in one point defect core
or grain
boundary are picked out because the director field changes drastically
on those sites. They are very near to each other. However, they may not be
neighbors. So we define a filter $a_0$, and
when any two points' distance is less than $a_0$, they are supposed to
be in the same defect or grain boundary's core region. We use the
cluster multiple labeling method of Hoshen and Kopelman
\cite{HK} to pick out
such point clusters. Thus given the
system's status at any time, we can find those sets of points 
corresponding to each individual defect or grain boundary.

Now we measure the approximate size of these structures and then
distinguish between point defects and grain boundaries. We use the number of
the points in the set as the size of the corresponding
structure. This approximation
reflects the actual size of the defect or grain
boundary quite well. Then we define a filter $l_0$, and when the structure's
size is larger than $l_0$, we regard the corresponding structure as a
grain boundary, otherwise it's taken to be a point defect (dislocation or
disclination).

We employed $a_0=5\,\Delta r$ where $\Delta r$ is the
lattice spacing and $l_0=18$. The results are quite satisfying.

After we have picked out the point defects, we can devide them into
disclinations and dislocations. We
follow Harrison's method \cite{Harrison,H2002}. Given the angle field
$\varphi({\bf x})$ computed in Eq.(A6) after the smoothing process, we do an integral of the
variation of $\varphi({\bf x})$ over
a counterclockwise close path around the ``mass-center'' of a point
defect, which is defined below. The condition for a defect to be a
disclination is 
\begin{equation}
\oint \frac{\partial \varphi}{\partial s}\,ds = \pm 2\pi
\end{equation}
The integral is zero if the defect is a dislocation. To make the
computation easier, we choose a $16\times 16$ square with the
mass-center at its center as the integration route.

To record the motion of one single defect or grain boundary, we track
the motion of the corresponding point set's ``mass center'', which is
defined as follows. Suppose the point set has $n$ points with
coordinates ${\bf r}_i,\,i=1,2,...,n$. Then the ``mass center'' of the
point set is defined as $\displaystyle {\bf r}=\sum_i {\bf r}_i/n$, just
like the usual mass center definition in classical mechanics but with
all masses equal to one.

In the evolution of the SH model, we sample the
system every $500$ time steps (in our case this is equal to 15
dimensionless time units), which is a quite short-time period in the 
simulation. We then identify all the dislocations, disclinations and grain boundaries,
distinguish among them, and compute their {\it centers~of~mass}.  
Suppose at time $t_1$, we have the set of {\it mass~ centers}
$P=\{{\bf p}_i,\, i=1,2,...,n_1\}$, and at time $t_2$, the ``mass center''
set is $Q=\{{\bf q}_j,\, j=1,2,...,n_2\}$; usually $n_1 \ne n_2$. Define
$d_{PQ}(i,j)=|\,{\bf p}_i-{\bf q}_j|$. We assume that the
defects and grain boundaries do not move much in such a short time
period. So if there exist two integers $k \in [1,n_1]$ and $l \in
[1,n_2]$, such that 
\begin{equation}
d_{PQ}(k,l)=\min_{j \in [1,n_2]}
d_{PQ}(k,j)=\min_{i \in [1,n_1]} d_{PQ}(i,l) \ ,
\end{equation}
it is quite 
reasonable to believe that ${\bf p}_k$ and ${\bf q}_l$ are just the same 
defect's or grain boundary's ``mass center'' at two successive times. Using 
this method, we are able to find out the trajectories of the ``mass
centers'' as time goes on. Not all points in $P$ and $Q$ can be grouped
into such pairs. On the one hand, this is because $n_1 \ne n_2$; on the
other hand, this is also due to the criterion (A9) applied onto
${\bf p}_k$ and ${\bf q}_l$. Physically, this is consistent with the
phenomena of the defect annihilation and the combination, split and shrinkage 
of
grain boundaries.

\section{Measurement of the Nematic Correlation Function}

To probe the stripes' increasingly orientational order, we define the
correlation function which is similar to the one employed by
Christensen and Bray\cite{CB}.
\begin{eqnarray}
\lefteqn{C_{nn}({\bf r},t)=\frac{1}{N^2} \sum_{{\bf x}} \langle \,
\cos\,[\varphi({\bf x}+{\bf r},t)-\varphi({\bf x},t)] \,\rangle {}} \nonumber 
\\
& & {}= \frac{1}{N^2} \sum_{{\bf x}} \langle \, \cos
\varphi({\bf x}+{\bf r},t)\cdot \cos \varphi({\bf x},t)\, \rangle + {}
\nonumber \\
& & {}+ \frac{1}{N^2}
\sum_{{\bf x}} \langle \, \sin
\varphi({\bf x}+{\bf r},t) \cdot \sin({\bf x},t) \,\rangle \ ,
\label{eq:26}
\end{eqnarray}
where $N^2$ is the area of the system and the angular brackets denote
the statistical average over different initial conditions. The
definition of the angle $\varphi({\bf x})$ is given in appendix A.

Now in Eq.(\ref{eq:26}) the function has been split into two parts 
which have the
same form
\begin{equation}
G({\bf r},t)=\frac{1}{N^2} \sum_{{\bf x}} \langle \,
f({\bf x}+{\bf r},t)\cdot f({\bf x},t) \,\rangle \ ,
\label{eq:27}
\end{equation}
with $f({\bf x},t)=\cos \varphi({\bf x},t)$ and $f({\bf x},t)= \sin
\varphi({\bf x},t)$ separately. Eq.(\ref{eq:27}) can be easily calculated by 
fast
Fourier transformation (FFT). First FFT  $f({\bf r},t)$\,
to obtain its Fourier components
$\tilde{f}({\bf k},t)$. Then $\tilde{G}({\bf k},t)= \langle \,
|\tilde{f}({\bf k},t)|^2 \, \rangle $, and inverse Fourier transformation
gives $G({\bf r},t)$. We
compute the two parts in Eq.(\ref{eq:27}) separately, and  then
add to obtain the
correlation function $C_{nn}({\bf r},t)$.

\end{multicols}

\begin{thebibliography}{99}
\bibitem{NT}  Assume that the total area of defects and grain
boundaries is $a(t)$, and $\langle\psi^2\rangle$ in those regions is a constant
$\chi^2$, which is smaller than $\langle \psi^2_0\rangle$. Then approximately
$\displaystyle \langle\psi^2\rangle_t
=\frac{1}{S}\,\left[\langle\psi_0^2\rangle\cdot\left(S-a(t)\right)+\chi^2\cdot
a(t)\right]=\langle\psi_0^2\rangle -
a\cdot\frac{\langle\psi_0^2\rangle-\chi^2}{S}$, i.e. $\Delta \psi^2(t)=
\langle\psi_0^2\rangle-\langle \psi^{2}\rangle_{t} \propto a(t)$.
Since the energy should
concentrate on the defects and grain boundaries, it is reasonable to
expect that $a(t) \propto \Delta
E(t) \propto L_E^{-1}(t)$, which is supported by the simulations. 
So $\Delta \psi^2(t)\propto
L_{E}^{-1}(t)$. $L_E(t)$ and $L_{\psi}(t)$ are expected to have the same
growth exponent. This is also supported by our simulations.
\bibitem{SHRINK}    One may doubt that the domain walls break up at this
time so that their sizes decrease. However if this was true, the number
of domain walls would increase, which is obviously wrong, as is shown in
Fig. 10. So the main effect that decreases the size of the domain walls
is the shrinkage.
\end{thebibliography}
\end{document}